\documentclass{aa}
\usepackage{graphicx}
\usepackage[varg]{txfonts}
\usepackage{hyperref}

\begin{document}
\title{3D hydrodynamic simulations of the Galactic supernova remnant CTB 109}

\author{J.~Bolte\inst{1}
\and M.~Sasaki\inst{2}
\and D.~Breitschwerdt\inst{1}} 

\offprints{J. Bolte, \email{bolte@astro.physik.tu-berlin.de}}

\institute{Zentrum für Astronomie und Astrophysik,
TU Berlin,
Hardenbergstraße 36,
10623 Berlin, Germany
\and
Institut für Astronomie und Astrophysik,
Universität Tübingen,
Sand 1,
72076 Tübingen, Germany
}


\abstract{
Using detailed 3D hydrodynamic simulations 
we study the nature of the Galactic supernova remnant (SNR) \object{CTB 109} (G109.1-1.0),
which is well known for its semicircular shape and a bright diffuse X-ray emission feature inside the SNR.
}{
Our model has been designed to explain the observed morphology, with a special emphasis on the bright emission
feature inside the SNR. Moreover, we determine the age of the remnant and compare our findings with X-ray observations.
With CTB~109 we test a new method of detailed numerical simulations of diffuse young objects, using realistic initial
conditions derived directly from observations.
}{
We performed numerical 3D simulations with the RAMSES code. The initial density structure has been directly taken from
$^{12}$CO emission data, adding an additional dense cloud, which, when it is shocked, causes the bright
emission feature.
}{
From parameter studies we obtained the position $(\ell , b)=(109.1545^\circ , -1.0078^\circ)$ for an elliptical
cloud with $n_\text{cloud}=25~\text{cm}^{-3}$ based on the preshock density from \emph{Chandra} data and a maximum diameter of 4.54~pc,
whose encounter with the supernova (SN) shock wave generates the bright X-ray emission inside the SNR. The calculated age of the remnant is
about 11,000~yr according to our simulations. In addition, we can also determine the most probable site of the SN explosion.
}{
Hydrodynamic simulations can reproduce the morphology and the observed size of the SNR CTB~109 remarkably well.
Moreover, the simulations show that it is very plausible that the bright X-ray emission inside the SNR is the
result of an elliptical dense cloud shocked by the SN explosion wave. We show that numerical simulations using
observational data for an initial model can produce meaningful results.
}
\maketitle

\section{Introduction}

In astronomy, complex morphologies of diffuse objects with filamentary or wispy structures, for example, are often
difficult to explain because the ambient medium is very inhomogeneous, and/or dynamical changes in the flow (e.g. shock waves)
modify the emission or absorption processes significantly.
If one would like to know if a certain feature is due to hydrodynamic interactions,
a numerical simulation could help to clarify the conditions under which it has originated. A typical case is the X-ray observation of 
supernova remnants (SNRs).
 Here we present a method that exploits the fact that the molecular
environment of an SNR has a high inertia, and does not evolve much within a typical early Sedov-Taylor phase time scale of the
order of $10^4$~yr. Therefore, the initial background model, if taken from $^{12}$CO observations, can be scanned
identically into a computational grid. However, the observational background is a projection onto the plane of
the sky and has to be extended to three dimensions. Nevertheless, such an initial model represents the real
situation much better than previous simplifications, such as a homogeneous background. Density inhomogeneities
lead to mass loading of the flow and hence to a quite different dynamical and thermal history of the plasma.

The Galactic SNR CTB 109 (G109.1-1.0) associated with the anomalous X-ray pulsar (AXP) \object{1E\,2259+586}
was first detected in X-rays by the satellite mission \emph{Einstein} in 1980 \citep{1980Natur.287..805G} and has been
regularly observed ever since. 

\begin{figure*}[t]
\centering
\makeatletter%
\if@twocolumn%
  \includegraphics[width=\columnwidth]{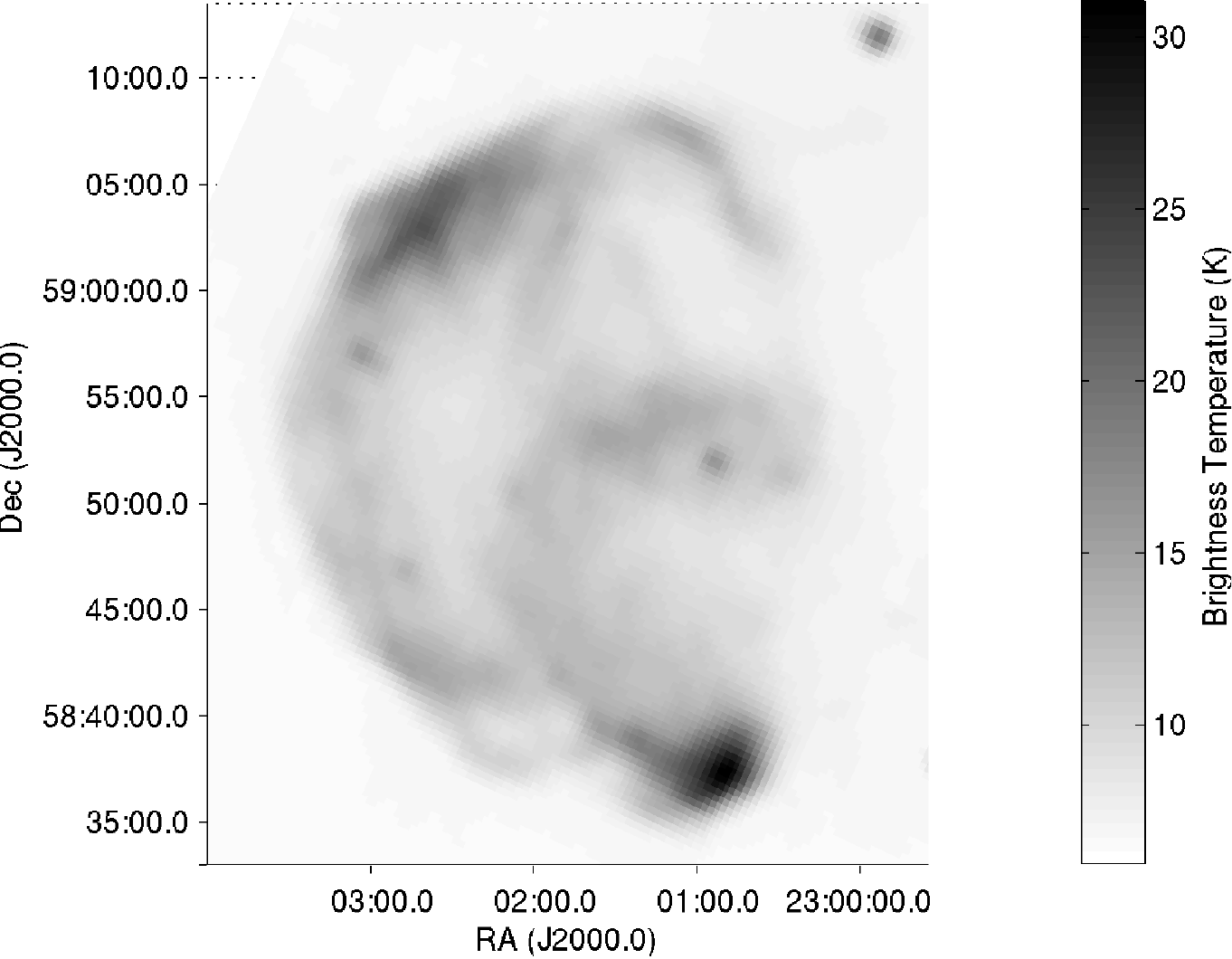}\hspace*{1cm}
  \includegraphics[width=0.8\columnwidth]{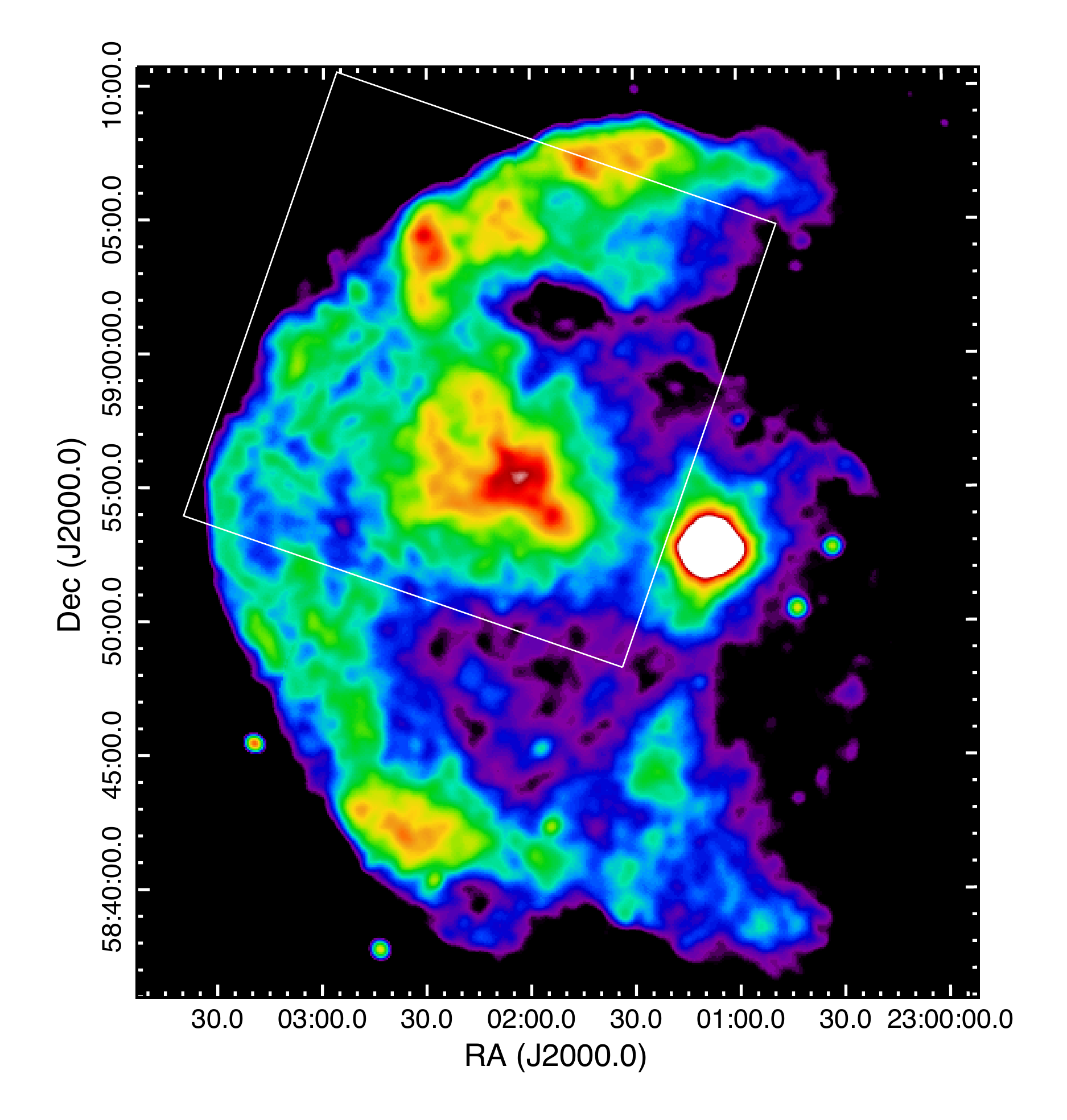}
\else
  \includegraphics[width=0.55\columnwidth]{ctb109_1420MHz.pdf}\hspace*{1mm}
  \includegraphics[width=0.45\columnwidth]{ctb109_xmm.pdf}
\fi
\makeatother
\caption{\emph{Left:} 1420-MHz radio continuum image of CTB~109 \citep{2002ApJ...576..169K} from the
Canadian Galactic Plane Survey \citep[CGPS,][]{2003AJ....125.3145T}.
\emph{Right:} Intensity map (0.3-4.0\,keV) of CTB~109 in false colour from the \emph{XMM-Newton} EPIC data \citep{2004ApJ...617..322S}.
The very bright point source is the AXP 1E\,2259+586 and the diffuse emission at RA $=$ 23\textsuperscript{h}02\textsuperscript{m},
Dec $=$ $+58^\circ55^\prime$ (J2000.0) with an extent of $\sim7^\prime$ is the Lobe. Chandra's field of view is marked as a white rectangle.}
\label{fig:xmm}
\end{figure*}

Radio observations at $\lambda 49~\text{cm}$ \citep{1981ApJ...246L.127H,1984ApJ...283..147H},
$\lambda 21~\text{cm}$ \citep{1984ApJ...283..147H,1986A&AS...63..345B}, $\lambda 4.6~\text{cm}$
\citep{1984ApJ...283..147H}, and at 2.7~GHz \citep{1983MNRAS.203..695D} show a morphology that corresponds well to that observed
in X-rays with the satellite \emph{Einstein}. However, no radio emission has been detected from the AXP. The analysis of CO observations
\citep{1987A&A...184..279T,2002ApJ...576..169K,2003AJ....125.3145T} has revealed that the western part of the SNR has
encountered a giant molecular cloud (GMC) complex, resulting in a semicircular shape of the remnant. Moreover, in radio a
double-shell structure can be seen (see Fig.~\ref{fig:xmm} (left)).
After the first detection with \emph{Einstein}, CTB~109 was observed in X-rays with \emph{ROSAT} PSPC and \emph{BBXT}
\citep{1993AAS...18310107R,1997ApJ...484..828R}, \emph{BeppoSAX} \citep{1998A&A...330..175P,1999NuPhS..69...96P}, and
\emph{ASCA} \citep{1998AdSpR..22.1039R}. Partial X-ray shells were identified in the east, south, and north surrounding the AXP, but no 
shell structure was found in the west owing to the GMC. Moreover, an X-ray bright interior region called the X-ray lobe
was identified in CTB~109. Higher resolution studies in X-rays were performed by \emph{XMM-Newton} \citep{2004ApJ...617..322S},
\emph{Chandra} \citep{2006ApJ...642L.149S,2013A&A...552A..45S}, and \emph{Suzaku} \citep{2015PASJ...67....9N} to investigate the remnant
in more detail. The $\gamma$-ray observations by \emph{Fermi}-LAT \citep{2012ApJ...756...88C} complement the multi-wavelength view
of CTB~109.

The determination of the distance to an SNR is quite challenging. Recent measurements show that the
distance to CTB~109 lies between 3 and 4~kpc \citep{2002ApJ...576..169K,2010MNRAS.404L...1T}; the latest
value is $3.2\pm 0.2~\text{kpc}$ \citep{2012ApJ...746L...4K}. 
It is even more difficult to determine the proper motion of the associated anomalous X-ray pulsar. The analysis
by \citet{2009AJ....137..354K} yields $(\mu_\alpha,\mu_\delta) = (-11\pm 20~\text{mas}\,\text{yr}^{-1},
-35\pm 20~\text{mas}\,\text{yr}^{-1})$, which is larger than the values $(\mu_\alpha,\mu_\delta) =
(-6.4\pm 0.6~\text{mas}\,\text{yr}^{-1}, -2.3\pm 0.6~\text{mas}\,\text{yr}^{-1})$ from \citet{2013ApJ...772...31T}
where the uncertainties are much smaller.

The X-ray observations with \emph{XMM-Newton} \citep{2004ApJ...617..322S} show two prominent features of the
SNR. Firstly, it has a semicircular shape due to a giant molecular cloud (GMC) complex in the west and secondly, a
bright diffuse emission region, which is known as the ``Lobe'', can be seen (see Fig.~\ref{fig:xmm} (right)).
It was discovered by \citet{1987A&A...184..279T} that the GMC complex extends even to the front of the remnant. Consequently,
the Lobe could simply result from a projection effect, i.e. the result of a ``hole'', a region with little or no
absorption for X-rays, in the GMC. Alternatively, this particular emission feature could be caused by a shock heated cloud
fragment, yielding bright X-ray emission. 

In \citet{2006ApJ...642L.149S} the velocity profiles (from high-resolution $^{12}$CO data from the Five College Radio Astronomy
Observatory) and \emph{Chandra} data of three molecular clouds around the Lobe were analysed. One of the clouds partly overlap
with the Lobe. In various parts of this cloud the $^{12}$CO velocity profile fits very well with a Gaussian. However, where the
Lobe and the cloud overlap, the velocity profile deviates from being Gaussian and has an additional component towards higher negative
velocities. This indicates an additional acceleration in this part of the cloud, e.g. by a supernova (SN) blast wave.
Moreover, the molecular hydrogen column densities in this region are relatively high, while the foreground absorption in X-rays is
lower than in other parts. This could also be explained by an interaction between a SN shock wave and a dense cloud. Consequently,
these results give new evidence for the hypothesis that the Lobe is the result of a shocked dense cloud.

In this work we will verify this hypothesis by 3D hydrodynamic simulations of the SNR. For this purpose an initial density
distribution, which is based on $^{12}$CO observations, is supplemented with an additional cloud. By varying the properties of this
homogeneous cloud, i.e. its position, size, shape, orientation, and density, the observed morphology and the X-ray emission of
CTB~109 can be reproduced very well.

\section{Hydrodynamic simulations}
\subsection{Previous modelling efforts}
\subsubsection{Analytic models}
The first efforts to model the evolution of CTB~109 were made by \citet{1981ApJ...246L.127H} who used the Sedov-Taylor
solution \citep{1959sdmm.book.....S,1946RSPSA.186..273T,1950RSPSA.201..159T} to estimate the age of the remnant.
For a homogeneous spherically symmetric medium they obtained an age of 17,000~yr from the \emph{Einstein} data. In 1991, when
the semicircular morphology of the SNR was already known, \citet{1990AcASn..31..121N} used the same method only for the
eastern part and obtained an age of 13,000~yr. For the western part they used the analytic solution for the pressure-driven snowplow phase
\citep[e.g.][]{1977ApJ...218..148M} to describe the propagation of the blast wave into the GMC. In \citet{2004ApJ...617..322S}
and \citet{2013A&A...552A..45S}, observational data again only from the eastern part obtained with \emph{XMM-Newton}
and \emph{Chandra}, respectively, were used to estimate the age of CTB~109 with the Sedov-Taylor solution. For the \emph{XMM-Newton}
data an age of $\sim 9,400\pm1,000~\text{yr}$ was derived while the \emph{Chandra} data yielded an age of
$14,000\pm2,000~\text{yr}$. The same age estimation was obtained from \emph{Suzaku} data by \citet{2015PASJ...67....9N} when applying the Sedov-Taylor
similarity solution. For the alternative assumption that the SNR is in the snowplow phase, they derive 10,000~yr as an age.

\subsubsection{Hydrodynamic models}
A first model of CTB~109 which is not based on a similarity solution is described by \citet{1992ApJ...388..127W}. They used
the thin shell approximation \citep{1960SPhD....5...46K} to describe the hydrodynamics of the semicircular SNR. Therefore, they
divided the system into a uniform interstellar medium in the east, in which a SN explodes, and a uniform dense molecular
cloud in the west. The interface between the two media is planar and 2~pc away from the explosion centre. Around the explosion centre,
the domain is divided into 100 rings for which the spherically symmetric hydrodynamic equations are solved, using a numerical code
by \citet{1988ApJ...324..776M} and \citet{1990ApJ...365..510C}. This simulation reproduces the semicircular morphology of CTB~109
quite well. For an initial energy of $E_0=3.6\cdot10^{50}~\text{erg}$, an ambient ISM density of
$n_0=0.13~\text{cm}^{-3}$, and a cloud density of $n_\text{cloud}=36~\text{cm}^{-3}$
the shape of the simulated remnant matches the \emph{Einstein} data after 13,000~yr.

In 2012 a sophisticated broadband study of CTB~109 was carried out by \citet{2012ApJ...756...88C}.  Their spherically
symmetric model includes hydrodynamics, efficient cosmic-ray acceleration, non-thermal emission, and a self-consistent calculation
of the X-ray thermal emission as described in detail by \citet{2007ApJ...661..879E} and \citet{2009ApJ...696.1956P}. The evolution
of the remnant with a power-law ejecta density profile is considered in a circumstellar medium (CSM) with uniform density. The main focus
of their study is to reproduce the broadband radio, X-ray, and $\gamma$-ray spectrum of CTB~109. For two distinct parameter
sets, describing the hadronic and leptonic components for the $\gamma$-ray flux reasonable fits were found.
However, because the X-ray spectrum of the SNR is dominated by thermal emission, a scenario where  both relativistic leptons and hadrons
contribute to the emission is more likely. Taking this into account, they found a better fit to the observations, which gives an ambient CSM density of
$n_0=0.5~\text{cm}^{-3}$ and an age of 11,000~yr for a distance to CTB~109 of 2.8~kpc.

\subsubsection{Comparison with our model}
Previous models of CTB~109 concentrated mainly on the estimation of the age \citep{1981ApJ...246L.127H,1990AcASn..31..121N,
2004ApJ...617..322S,2013A&A...552A..45S,2015PASJ...67....9N}, only \citet{1992ApJ...388..127W} aimed at reproducing of a simple semicircular
morphology, and \citet{2012ApJ...756...88C} focused on the broadband emission. For this purpose, simple spherically symmetric hydrodynamic models
are sufficient. However, aiming at a detailed simulation of the observed complex morphology, which differs from being purely
semicircular, no spherically symmetric model can be used. Therefore, we use a highly inhomogeneous ambient density structure from observation,
which includes the GMC in the west. This would favour a 2D simulation. However, developing turbulent structures and the occurring instabilities,
which yield the break-up of the SN shell, are entirely different in 2D than in 3D \citep{2004ARA&A..42..211E,2000ApJ...528..989K}.
While turbulence in nature is an inherently 3D process, one might think of turbulent flows that are approximately 2D, in the sense that large scale
coupling of eddies clearly exceeds the extension in one particular dimension. However, the physical properties between 2D and 3D turbulence are vastly
different, so that sacrificing one dimension for a higher spatial resolution might lead to erroneous results. In contrast to 3D, as has been noted by
\citet{1969PhFl...12..II233B} and \citet{1967PhFl...10...1417K}, due to the conservation of vorticity in 2D (in the incompressible, inviscid
case without external forcing), the energy cascade is reversed, i.e. from small to large scales, while enstrophy flows in
the opposite direction (double-cascade).
In particular, we want to model the bright X-ray emission feature inside the SNR for the first time as a result of a shocked dense cloud. For
this reason an additional cloud is added to the initial density structure, precluding again a 1D or 2D model.

To estimate the age of CTB~109 and to reproduce its semicircular morphology, a complex 3D model is not necessary and make it even more difficult
to study many different parameter sets. However, our simulations complement the previous studies on CTB~109 by analysing the effect of an
inhomogeneous medium on the evolution of the SNR and by simulating the bright X-ray emission region inside the SNR by a shocked dense cloud.
These aspects could not be studied with their spherically symmetric models.

The details of the thermal and non-thermal emission of the SNR, which are modelled in detail by \citet{2012ApJ...756...88C}, are beyond
the scope of our study. We investigate in detail the X-ray emissivity to compare our finding with the X-ray observations. In forthcoming studies,
spectral modelling will also be considered.

\subsection{Numerical details of our studies}
Reproducing observations by  numerical simulations represents an inverse
problem, which, as it is well-known, does not always possess a unique solution or even
a solution at all. We simplify the problem by assuming that the background medium, due to its high inertia, has
not changed significantly during the short evolution of the SNR, or in other words, we use the
present ambient medium as the initial model.

The numerical simulations were performed with the RAMSES code \citep{2002A&A...385..337T}. The code works on 1D, 2D, or 3D
unstructured Cartesian grids and solves the discretised Euler equations in their conservative form with a second-order Godunov scheme
for a perfect gas (Piecewise Linear Method).
In particular, the MUSCL scheme \citep[Monotonic Upstream-Centred Scheme for Conservation Law,][]{1976cppa.conf...E1V,vanLeer1979101}
with the HLLC Riemann solver \citep{1994ShWav...4...25T} was used where the MinMod slope limiter \citep{Roe1981} is employed to
obtain a total variation diminishing discretisation scheme \citep{1983JCoPh..49..357H}. A 58~pc cube, which comprises the SNR
and the GMC, is initially uniformly discretised into $128^3$ cells, hence the initial resolution of the grid
is 0.45~pc. Based on a refinement strategy, which refines cells adaptively if steep pressure gradients occur, the grid is
refined during the evolution of the SNR. The finest resolved structures due to the adaptive mesh refinement (AMR) have a size of 0.1~pc.

Given that the age of the SNR is $\sim 10^4~\text{yr}$, we can omit its free-expansion phase in the
simulation because the duration of this phase is only $\lesssim 1000~\text{yr}$. The reverse shock must have
formed and already reached the centre of the SNR heating the ejecta. Therefore, the Sedov-Taylor similarity solution can be applied.
Even though the ejecta might have had inhomogeneities like a Si clump as suggested by \citet{2013A&A...552A..45S}, such clumps
should have no significant effect on the overall evolution of the forward shock of the SNR expanding into the ISM. Furthermore, the
end of the free-expansion phase is characterised by the fact that the mass of the swept-up interstellar material exceeds the mass of the
ejecta, i.e. we can neglect the ejecta in our simulations. Consequently, the SN explosion can be modelled by a point-explosion, i.e. the
pressure $p_0$ in  the $2^3$ cells at the explosion centre is given as $p_0=3\,(\gamma-1)\,E_0/(4\,\pi\,(\Delta x)^3)$ with
$\Delta x=0.45~\text{pc}$ which yields the expanding remnant. As long as the dynamical time $t$ is smaller than the cooling time
\begin{equation}
 t_\text{cool} = \frac{3\, k_\text{B}\, T}{2\, n\, \Lambda(T,n)}
 \label{eq:1}
\end{equation}
\citep[e.g.][]{1978ppim.book.....S}
with number density $n$, temperature $T$, cooling function $\Lambda$, and Boltzmann constant $k_\text{B}$, the expansion is adiabatic.
Because the effects of cooling and heating cannot be neglected in interaction regions
with dense matter, a cooling function, which was computed with the freely available spectral synthesis code CLOUDY \citep{1998PASP..110..761F},
is employed in the RAMSES code.

The CLOUDY code simulates the conditions within an astrophysical plasma and then predicts the emitted spectrum. In particular,
the ionisation level in the plasma is determined by balancing all ionisation processes (photo, Auger, and collisional ionisation as well as
charge transfer) and all recombination processes (radiative, low-temperature dielectronic, high-temperature dielectronic, three-body
recombination, and charge transfer) of the 30 lightest elements H -- Zn in all stages. For this purpose, an optically thick gas slab is
divided into a large number of zones, such that constant conditions can be assumed in each zone. For the free electrons in each
zone it is assumed that their velocity is predominately Maxwellian distributed with a kinetic temperature determined by the
balance between heating (e.g. photoelectric, mechanical, cosmic-ray) and cooling (e.g. forbidden-line cooling and inelastic collisions
between electrons and other particles). Simultaneously, the associated line and continuum radiative transfer is solved. The resulting cooling function
for solar abundances, dependent on the number density and the temperature, is then used by the RAMSES cooling routine.

\subsection{Model setup}
For a realistic 3D numerical simulation of a supernova explosion, which reproduces a geometry comparable to that of CTB~109, we use
the $^{12}$CO emission of the Canadian Galactic Plane Survey \citep[CGPS,][]{2003AJ....125.3145T} at a radial velocity of
$v=-51~\text{km}~\text{s}^{-1}$ as an indicator of the inhomogeneous ambient density structure. This reflects the fact that the
molecular material associated with CTB~109 has approximately this radial velocity \citep{1990ApJ...351..157T}. The idea of using this as
a density indicator is based on the assumption that the GMC complex in the west was hardly displaced by the shock wave of the
explosion (initial explosion energy $E_0= 10^{51}~\text{erg}$) because of its high inertia.

For the conversion of the $^{12}$CO data to an ambient density structure, results from \emph{XMM-Newton}
\citep{2004ApJ...617..322S} and \emph{Chandra} observations \citep{2013A&A...552A..45S}, respectively, are used. For this purpose,
the mean $^{12}$CO emission is associated in each case with the derived preshock density value, yielding a conversion factor
to convert the $^{12}$CO emission to a density structure. The \emph{XMM-Newton}
observations aimed at the studies of the morphology CTB~109 and any morphological connection between the AXP and the Lobe, which was
not found. Moreover, only little spectral variation across the remnant were found. The obtained spectra of the shell and the Lobe
could be well fitted with a one-component non-equilibrium ionisation (NEI) model, describing a mixture of the shocked ISM and shocked
ejecta. The preshock density of the ambient medium was derived to be $n_0 = (0.16\pm 0.02)\, d_3^{-0.5}~\text{cm}^{-3}$, where $d_3 = D/3.0~\text{kpc}$
is a scaling factor accounting for the distance $D$ of CTB~109, which was estimated as $3.2\pm 0.2~\text{kpc}$ by
\citet{2012ApJ...746L...4K} with the new method of \citet{2006ApJ...644..214F}. This results in a density of the medium of $n_0 = 0.155~\text{cm}^{-3}$
in case of the \emph{XMM-Newton} data.

The study by \citet{2013A&A...552A..45S} of the \emph{Chandra} data is complementary to the \emph{XMM-Newton}
study. Only the Lobe and the north-eastern part of CTB~109 were observed to study the Lobe in greater detail.
In comparison to the \emph{XMM-Newton} observations, more point sources were found and structures, for example in the Lobe,
could be resolved. For the spectral analysis the emission region was divided into small regions with similar
surface brightness resulting in a higher spatial resolution than for the \emph{XMM-Newton} observations. In 50\%
of the small regions a two-component NEI model, describing the shocked ISM and the shocked ejecta separately,
led to significantly better fits than a one-component NEI model. For the density of the ISM a median value
$n_{1,\text{median}}=1.2~\text{cm}^{-3}$ was found. Assuming a compression ratio of 4 ($\gamma=5/3$), this
yields  $n_0 = 0.3~\text{cm}^{-3}$ as preshock density in case of the \emph{Chandra} data.

\begin{figure}[t]
\centering
\includegraphics[width=\columnwidth]{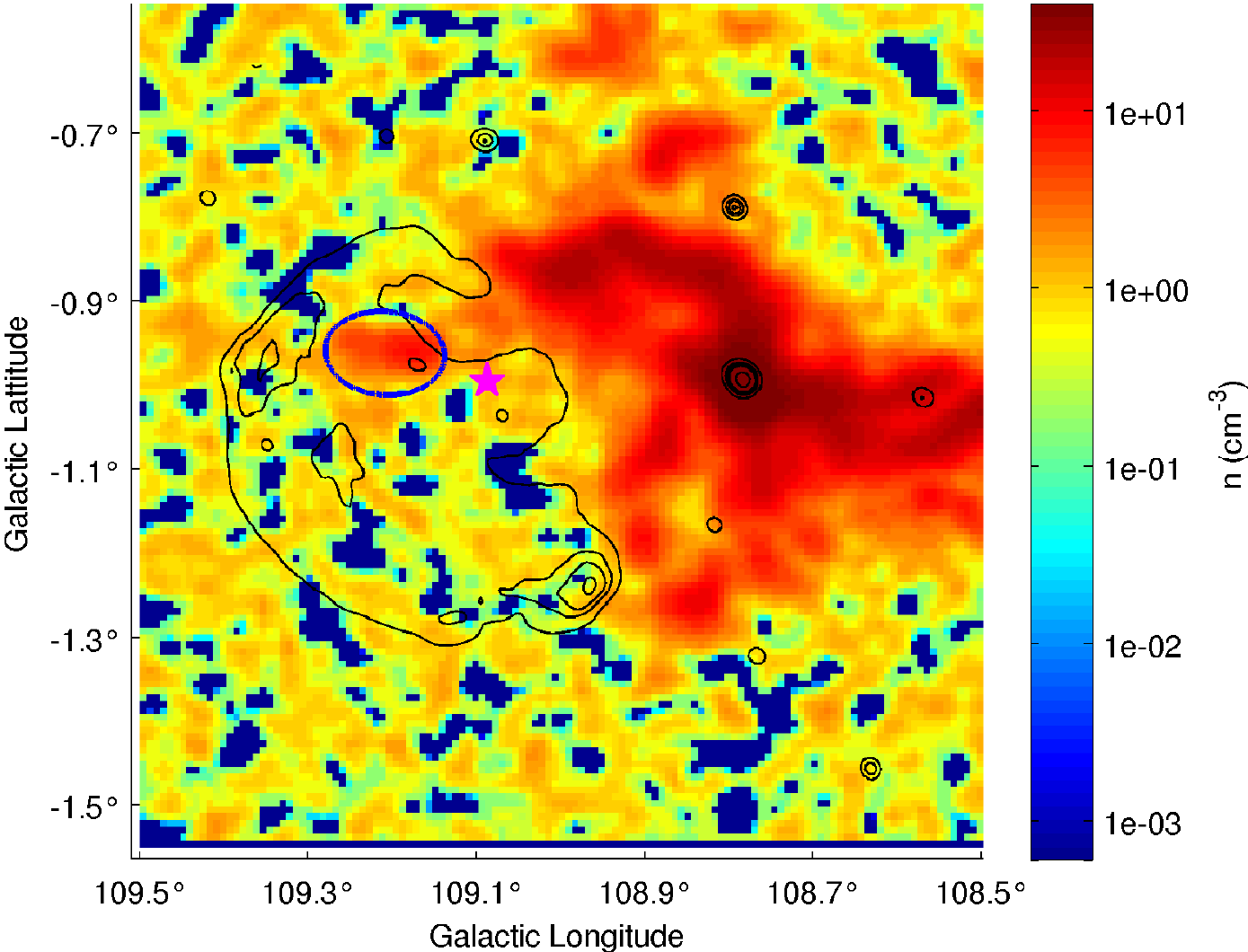}
\caption{Density structure of CTB~109, obtained by converting the mean $^{12}$CO emission 
in the eastern part with $n_0 = 0.155~\text{cm}^{-3}$ from \emph{XMM-Newton} data. Overlaid black contour lines display
the 1420-MHz continuum emission for 8~K, 15~K, 21~K, 29~K,  and 100~K \citep{2010MNRAS.404L...1T}.
The current position of the associated AXP 1E\,2259+586 is marked with a magenta star.
The circumference of a foreground cloud fragment, which was removed for the calculations, is marked as a blue ellipse.}\label{fig:co_into_density}
\end{figure}
We have performed our simulations for both possibilities of the ambient medium, i.e. for a mixture of
ISM and ejecta as a single fluid with $n_0 = 0.155~\text{cm}^{-3}$ (\emph{XMM-Newton}) and only for the ISM
with $n_0 = 0.3~\text{cm}^{-3}$ (\emph{Chandra}), respectively. As we are aiming to explain the Lobe as a shocked
dense cloud, the shape, size, orientation, and position of this Lobe-forming cloud should be independent from
the assumption of the mean preshock density.

Both density values were used separately to convert the $^{12}$CO emission of the CGPS into a density structure by assuming that a
typical $n_0$ corresponds to the mean value of the $^{12}$CO emission in the eastern part (see Fig.~\ref{fig:co_into_density}).

For the hydrodynamic simulations the 3D initial density distribution is modelled as a cylinder
with the long axis of the cylinder running along the line of sight, while its cross section is shown in
Fig.~\ref{fig:co_into_density}.
Initially, for the thermal pressure a uniform temperature value of 20~K is adopted and the
entire structure is considered to be at rest. 

We tested the hypothesis that the Lobe is the result of an encounter of a shock wave with a dense cloud.
A simply shaped cloud (resembling the observed Lobe region \emph{after} its encounter with the shock wave) 
is introduced ad hoc into the initial density distribution.

\section{Results}
\subsection{Derived initial properties of the Lobe-forming cloud}

\begin{table}[t]
\caption{Variation ranges of the parameter studies to determine the initial properties of the cloud, which develops into
the Lobe.}\label{tab:parameterstudies}
\centering
\begin{tabular}{ccc}
\hline\hline
Varied quantity & Range & Minimum step size\\\hline
position $\ell$	& $109.2^\circ$ -- $109.1^\circ$   & $0.01^\circ$\\
position $b$	& $(-1.1^\circ)$ -- $(-0.9^\circ)$  & $0.01^\circ$\\
maximum size	& 0.45-8.1~pc  & 0.45~pc\\
shape		& spherical, elliptical  & \\
orientation	& $(-90^\circ)$ -- $(+90^\circ)$  & $5^\circ$\\
density		& $1-60~\text{cm}^{-3}$  & $0.5~\text{cm}^{-3}$\\\hline
\end{tabular}
\end{table}

\begin{figure}[t]
\centering
\includegraphics[width=\columnwidth]{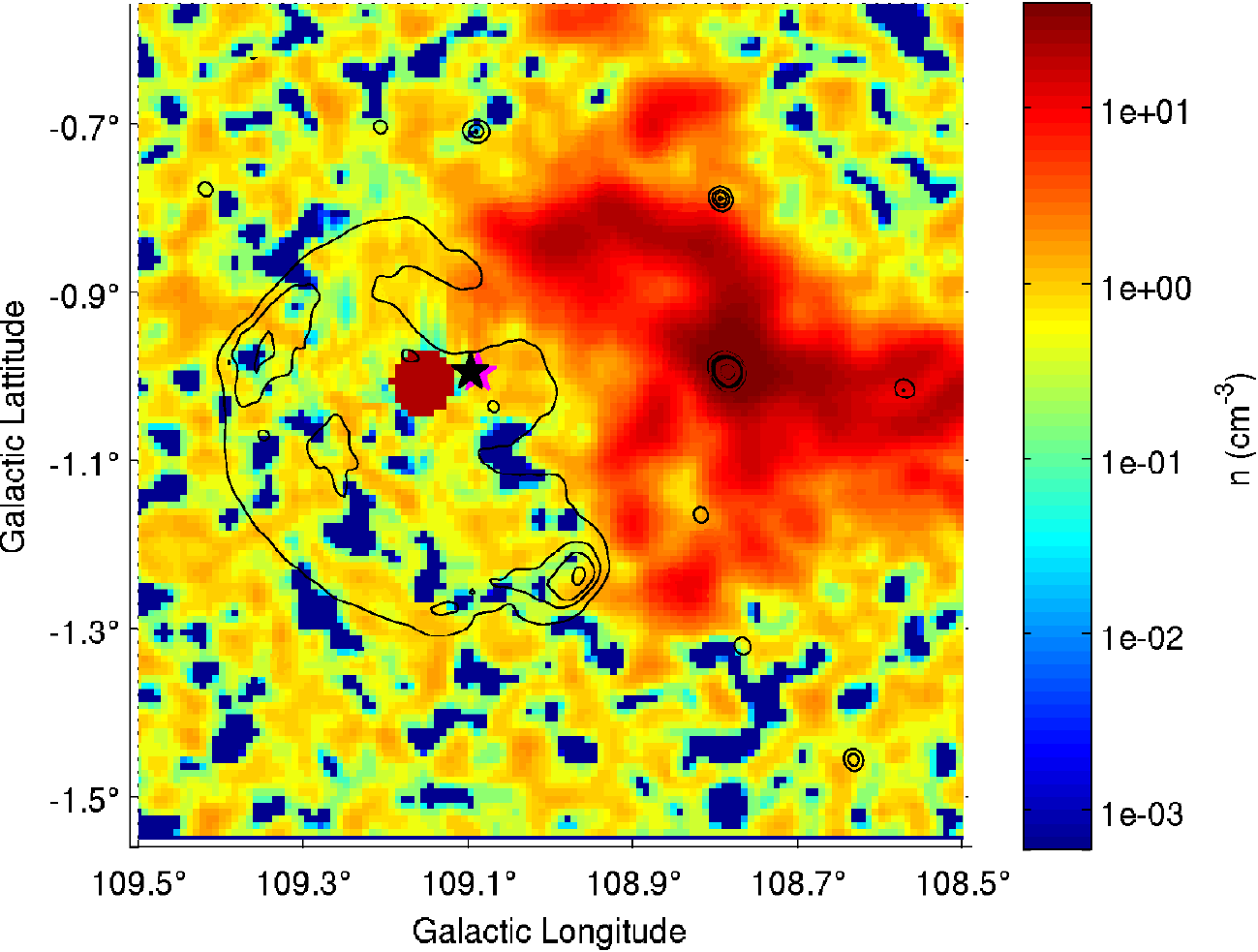}
\caption{Initial density structure from \emph{XMM-Newton} data, which gives the best fit of the 1420-MHz contours of CTB~109
and the Lobe region after 8,000~yr. The position of the supernova explosion is marked with a black star and the position
of the AXP with a magenta star, which is almost completely covered by the black star. The elliptically shaped cloud ($n_\text{cloud} = 20~\text{cm}^{-3}$), which
resembles the Lobe region after its encounter with the shock wave,
is located at $(\ell , b)=(109.1545^\circ , -1.0078^\circ)$.}\label{fig:initial}
\end{figure}

Based on the observed shape of the Lobe and the morphology of the remnant a good fit for the initial
density models from \emph{XMM-Newton} and \emph{Chandra} data, respectively, could be obtained
(see Fig.~\ref{fig:initial} for the model from \emph{XMM-Newton} data) by varying the position, size,
shape, orientation, and density of a simply shaped cloud. Especially, the position, size,
and density are very sensitive parameters, which significantly influence the results. Therefore, we put
special emphasis on determining these quantities. Over 800 different model configurations were tested
for both mean preshock densities, for details on the parameter studies see Table~\ref{tab:parameterstudies}.
Moreover, the original position of the supernova explosion is derived from the remnant age obtained from the simulation
and the proper motion of the AXP 1E\,2259+586, which is associated with the SNR CTB~109. In
\citet{2013ApJ...772...31T} its proper motion was derived as
$(\mu_\alpha,\mu_\delta) = (-6.4\pm 0.6~\text{mas}\,\text{yr}^{-1}, -2.3\pm 0.6~\text{mas}\,\text{yr}^{-1})$.

For $n_0 = 0.155~\text{cm}^{-3}$ (\emph{XMM-Newton} data) our calculation, based on the cloud properties shown in Fig.~\ref{fig:initial},
yields a remnant age of 8,000~yr, which is resulting in the most probable site of the explosion is $(\ell , b)=(109.0962^\circ , -0.9942^\circ)$, also marked
in Fig.~\ref{fig:initial}.

For the preshock density value from \emph{Chandra} observations ($n_0 = 0.3~\text{cm}^{-3}$) our calculation
yields a remnant age of 11,000~yr and hence the most probable site of the explosion is $(\ell , b)=(109.0995^\circ , -0.9936^\circ)$.

For both preshock density values the dense homogeneous cloud, which becomes the Lobe region, is modelled as an elliptical
cylinder of height 1.81~pc at $(\ell , b)=(109.1545^\circ , -1.0078^\circ)$ with a semi-major axis of 2.27~pc and a semi-minor axis
of 2.04~pc, which is oriented $5^\circ$ towards the east. In the case of the \emph{XMM-Newton} data the cloud has a density
of $n_\text{cloud} = 20~\text{cm}^{-3}$ while for the \emph{Chandra} data the density is $n_\text{cloud} = 25~\text{cm}^{-3}$.
We note that the modelled final structure of the  X-ray feature depends quite
sensitively on the initial properties of the Lobe-forming cloud, while the morphology of the remnant remains almost
unaffected. For instance, a higher density leads to a ring-like emission region, whereas  a lower density leads only to a
diffuse emission. Moreover, the derived age of the SNR also depends on the detailed properties of the cloud; for example
a cloud which is cut in half, but has the same total mass as before, yields an age of the SNR of $\sim10,000~\text{yr}$
for $n_0 = 0.155~\text{cm}^{-3}$ because the evolution of the feature has to be followed for a longer time to give good agreement
with the \textsc{Hi} data.

\subsection{Resulting structures of the SNR}
In the following we discuss simulations for both the \emph{XMM-Newton} and \emph{Chandra} data in order to assess the variations of
the model parameters compatible with existing observations.

\subsubsection{Estimation of the age}

\begin{figure*}[t]
\centering
\makeatletter%
\if@twocolumn%
  \includegraphics[width=\columnwidth]{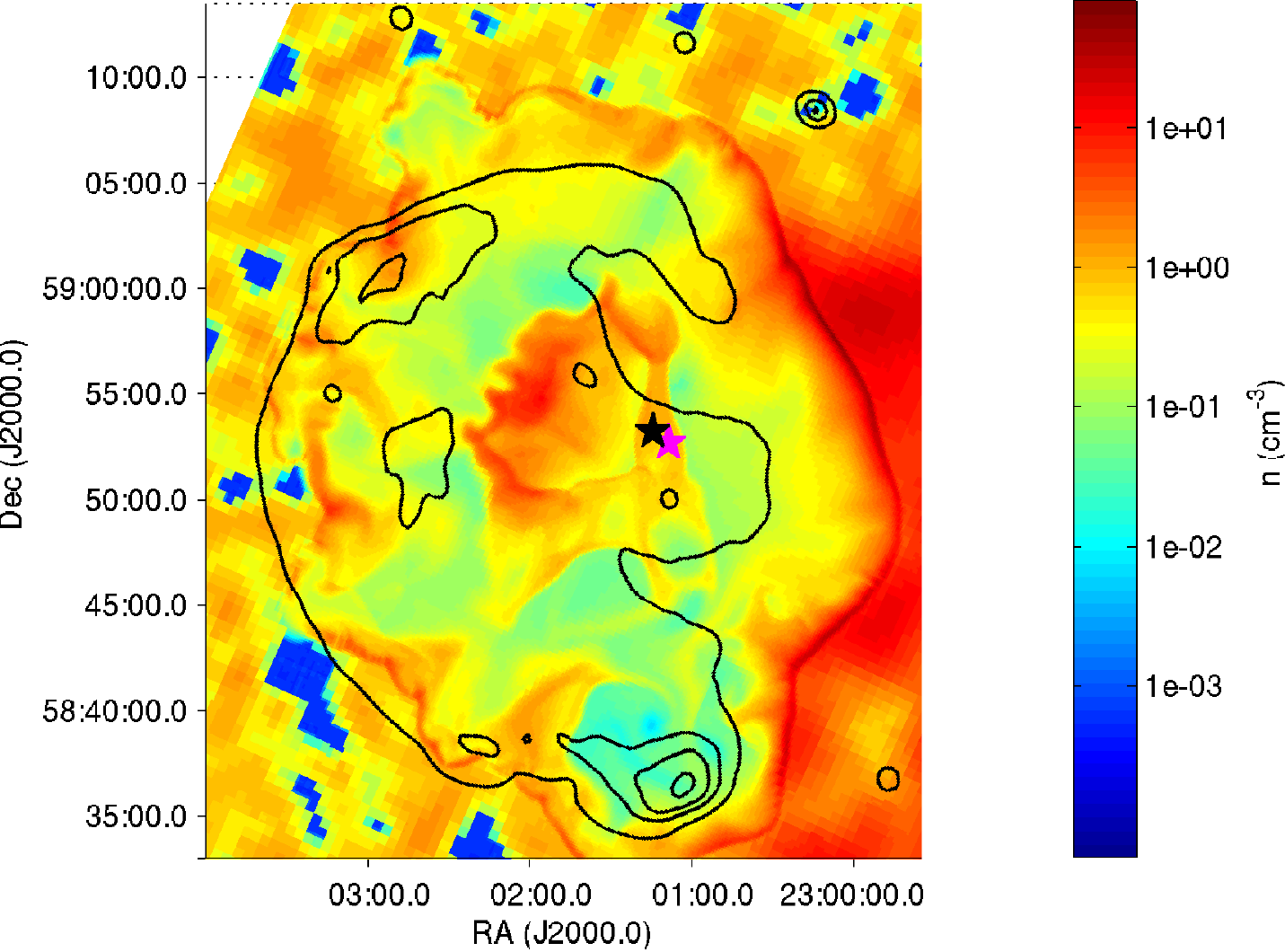}\hspace*{2mm}
  \includegraphics[width=\columnwidth]{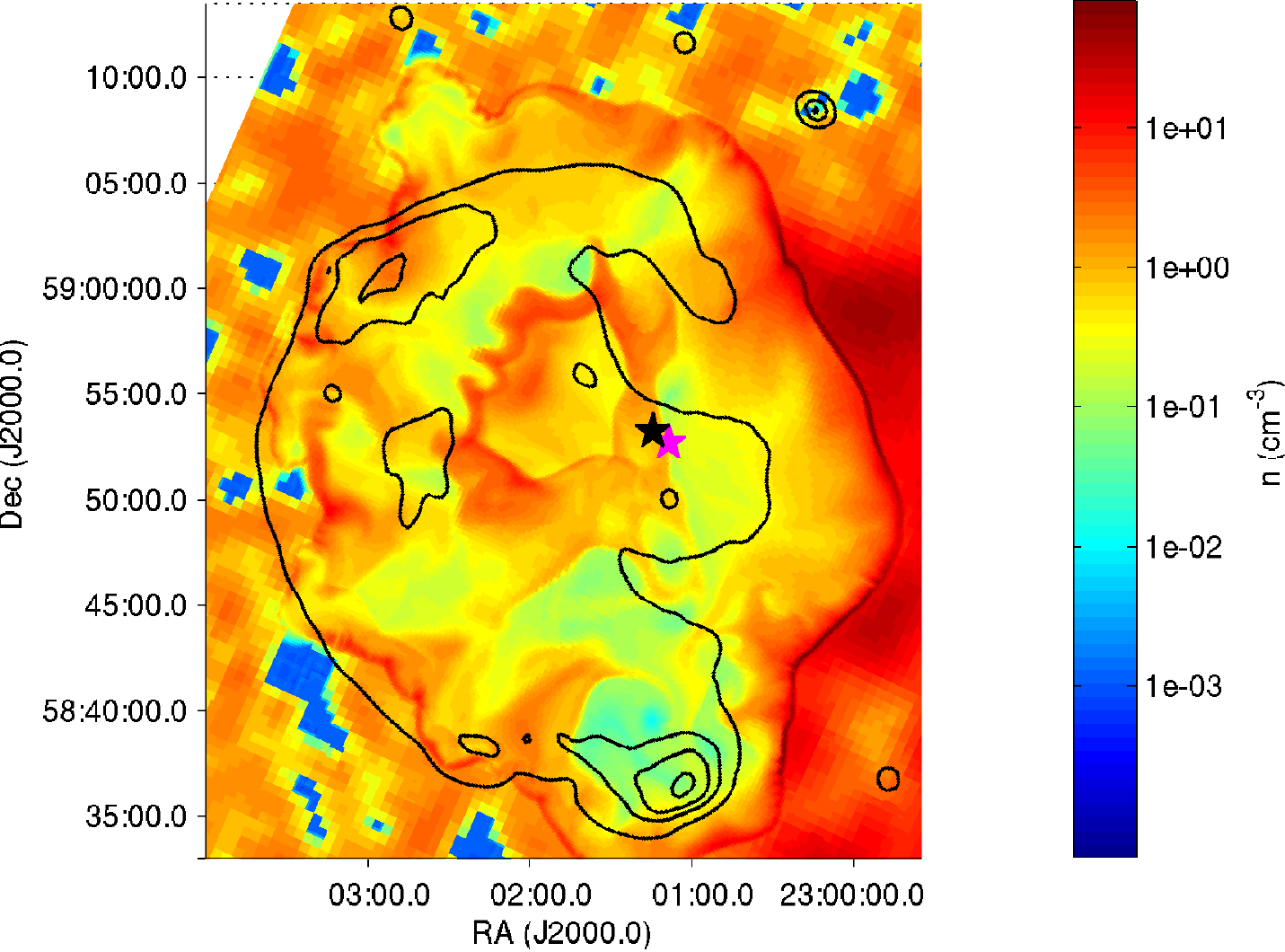}
\else
  \includegraphics[width=0.49\columnwidth]{density.pdf}
  \includegraphics[width=0.49\columnwidth]{density_chandra.pdf}
\fi
\makeatother
\caption{{\emph{Left:}} Resulting density distribution from the model fit after 8,000~yr for the \emph{XMM-Newton} data with
the 1420-MHz contours of CTB~109 overlaid in black. 
{\emph{Right:}}
Resulting density distribution from the model fit after 11,000~yr for the \emph{Chandra} data with
the 1420-MHz contours of CTB~109 overlaid in black. 
The position of the supernova explosion is marked with
a black star and the position of the AXP with a magenta star.}
\label{fig:density}
\end{figure*}

Assuming the preshock density distribution derived from the
\emph{XMM-Newton} data and a supernova that occurred
at $(\ell , b)=(109.0962^\circ , -0.9942^\circ)$, 
the resulting density structure after 8,000~yr matches very well the 1420-MHz 
contours of CTB~109 in the eastern and southern region
as illustrated in Fig.~\ref{fig:density} (left).
For the increased preshock density from the \emph{Chandra} data and the site of the explosion at $(\ell , b)=(109.0995^\circ , -0.9936^\circ)$,
the resulting density structure after 11,000~yr (see Fig.~\ref{fig:density},
right) matches again very well the 1420-MHz contours of 
CTB~109 in the eastern and southern region.

In summary, we can infer that the evolution time of the SNR is in general much smaller than the cooling time for both models,
such that SNR CTB~109 is still in its Sedov-Taylor phase. However, this is of course not the case for the Lobe-forming cloud
and the interaction region with the GMC where cooling cannot be neglected. Given the cooling time as in Equation~\eqref{eq:1},
we see that after $t \sim 1,000~\text{yr}$, when the shock has overrun the  Lobe-forming cloud and has heated it up, the ratio $t/t_\text{cool}$
becomes greater than 1. The cooling time in the cloud remains thereafter at $\sim 1,000~\text{yr}$. The same happens after $t \sim 4,000~\text{yr}$
in the western part where the GMC is. There the cooling time goes even down to $\sim 100~\text{yr}$. In all other parts of the SNR
$t_\text{cool}$ varies between $\sim 10^5~\text{yr}$ and $\gtrsim 10^6~\text{yr}$, such that the expansion into the eastern ambient
medium is adiabatical. We note that simulations of the remnant without cooling gave almost identical results for CTB~109, which supports our conclusion
that the SNR is in its Sedov-Taylor phase.

\subsubsection{Computed emissivities}
To compare the numerical simulations with the X-ray intensity from \emph{XMM-Newton} observations \citep{2004ApJ...617..322S},
we computed the X-ray emissivity $\varepsilon=n^2 \, \Lambda(T,n)$ from both simulations, where $\Lambda$ denotes the
X-ray cooling function from CLOUDY.

\begin{figure*}[t]
\centering
\makeatletter%
\if@twocolumn%
  \includegraphics[width=\columnwidth]{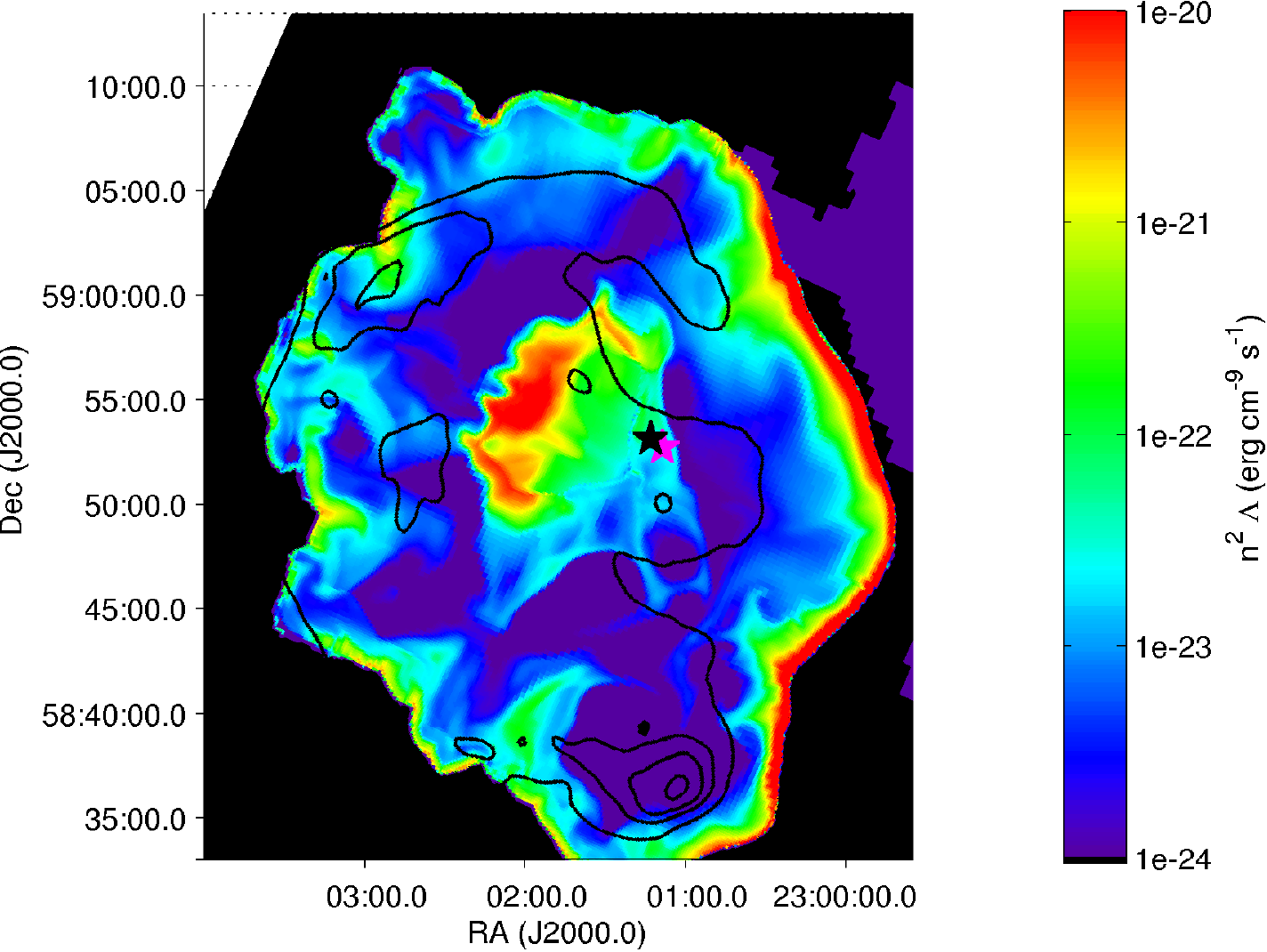}\hspace*{2mm}
  \includegraphics[width=\columnwidth]{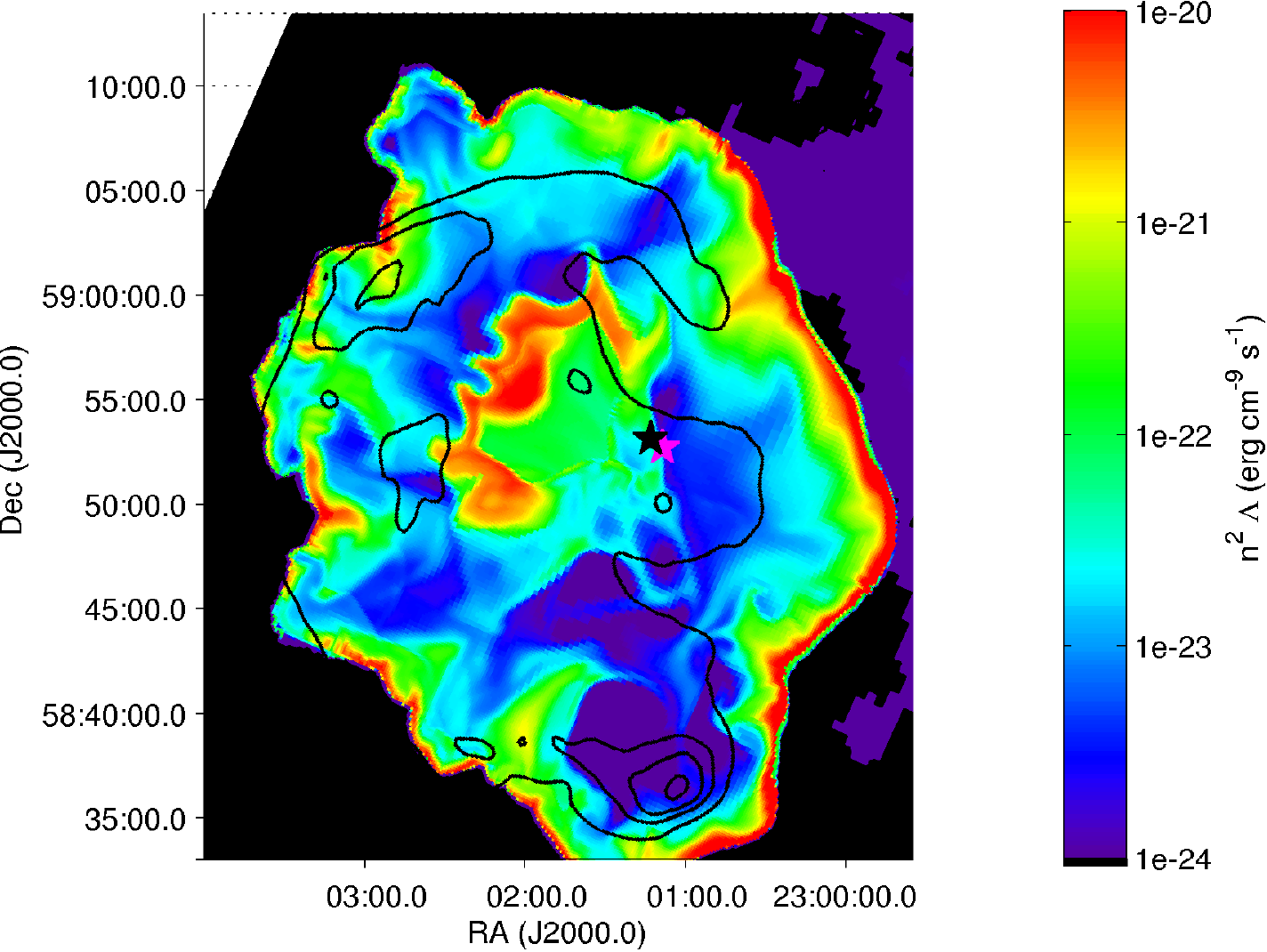}
\else
  \includegraphics[width=0.49\columnwidth]{emissivity.pdf}
  \includegraphics[width=0.49\columnwidth]{emissivity_chandra.pdf}
\fi
\makeatother
\caption{{\emph{Left:}}
Modelled X-ray emissivity $\varepsilon=n^2 \, \Lambda(T,n)$ after 8,000~yr (\emph{XMM-Newton} density model) simulating the intensity
map (0.3-4.0\,keV) of CTB~109 from the \emph{XMM-Newton} EPIC data \citep{2004ApJ...617..322S} (cf. Fig.~\ref{fig:xmm} (right)).  
{\emph{Right:}}
Modelled X-ray emissivity $\varepsilon=n^2 \, \Lambda(T,n)$ after 11,000~yr (\emph{Chandra} density model)
using the same intensity map (0.3-4.0\,keV) of CTB~109 from the \emph{XMM-Newton} EPIC data.
The position of the supernova explosion is marked with
a black star and the position of the AXP with a magenta star.}\label{fig:emissivity}
\end{figure*}

In Fig.~\ref{fig:emissivity} (left)
the simulated X-ray intensity for the preshock density obtained
from \emph{XMM-Newton} data is displayed.
The simulated X-ray intensity for the \emph{Chandra} density model is shown in Fig.~\ref{fig:emissivity} (right).

It can be seen by comparison with Fig.~\ref{fig:xmm} (right) that in both models the bright X-ray emission in the Lobe region can be reproduced very well,
while its morphology is only poorly matched by our model assumption of a shocked and simply shaped additional cloud. For a better agreement with
the observed Lobe structure the additional cloud needs to be inhomogeneous and of complex shape resulting in a very large number
of free parameters, which is beyond the scope of this study.

\subsubsection{Computed velocities}

\begin{figure*}[t]
\centering
\makeatletter%
\if@twocolumn%
  \includegraphics[width=\columnwidth]{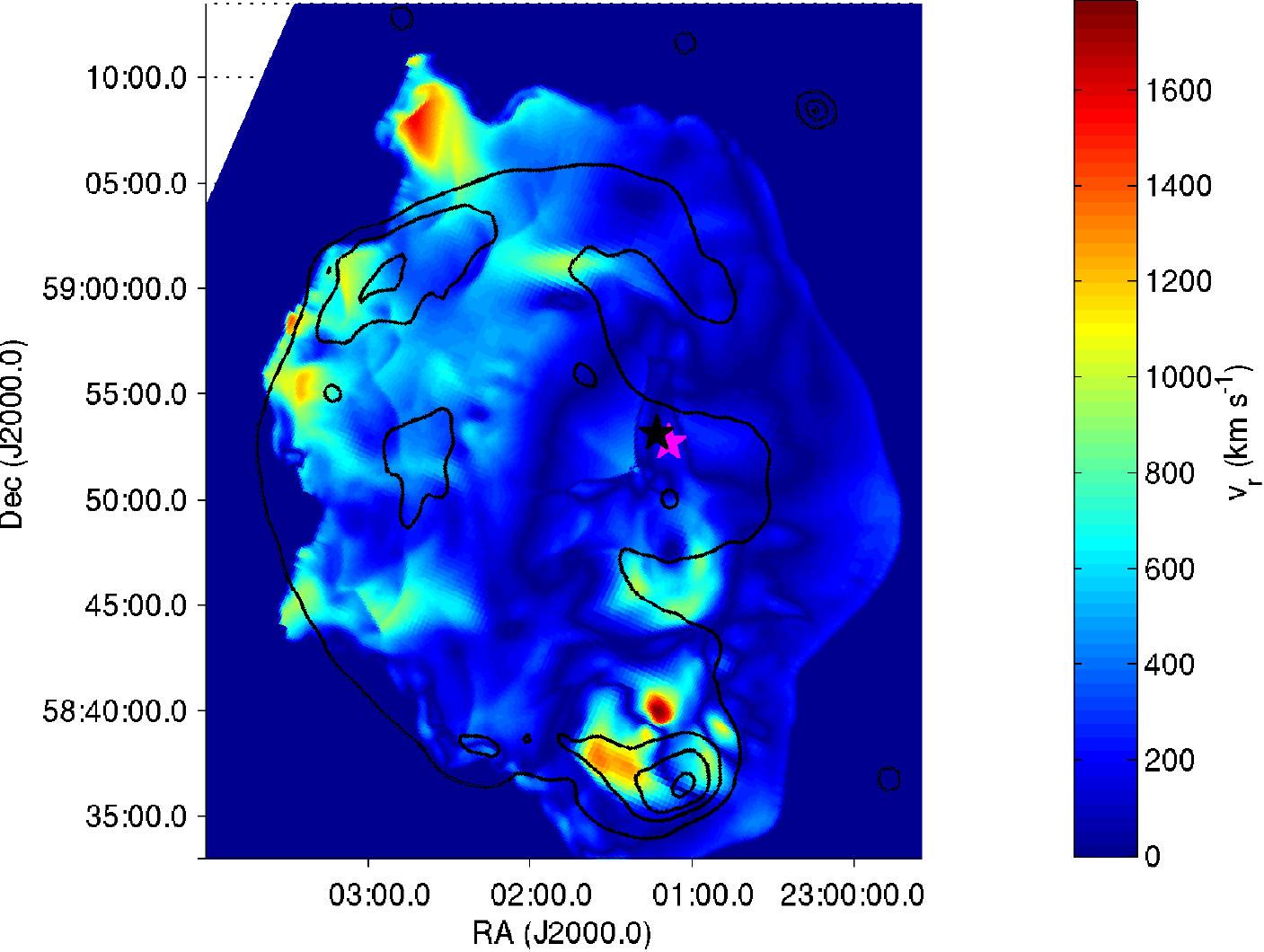}\hspace*{2mm}
  \includegraphics[width=\columnwidth]{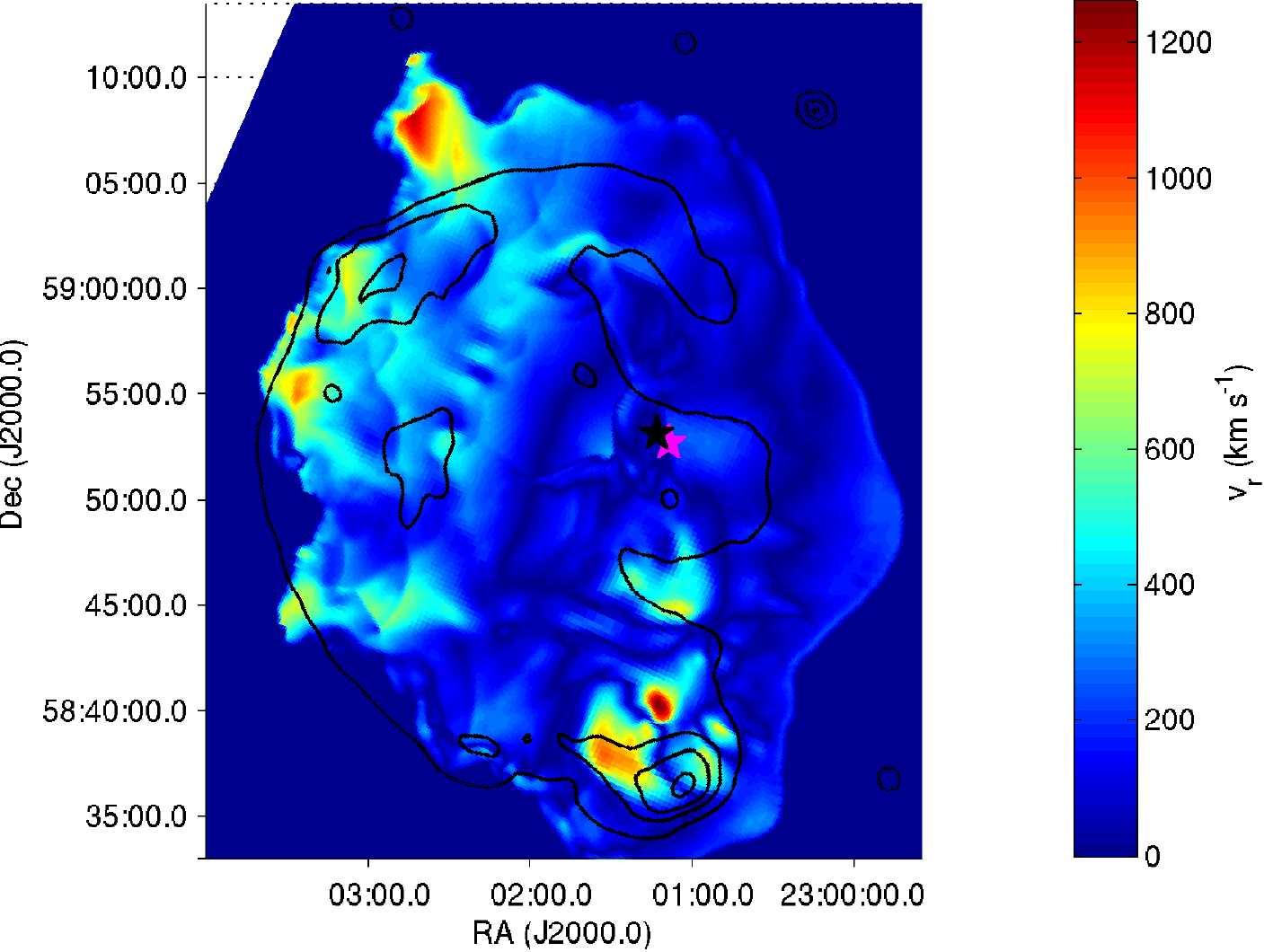}\\
  \includegraphics[width=\columnwidth]{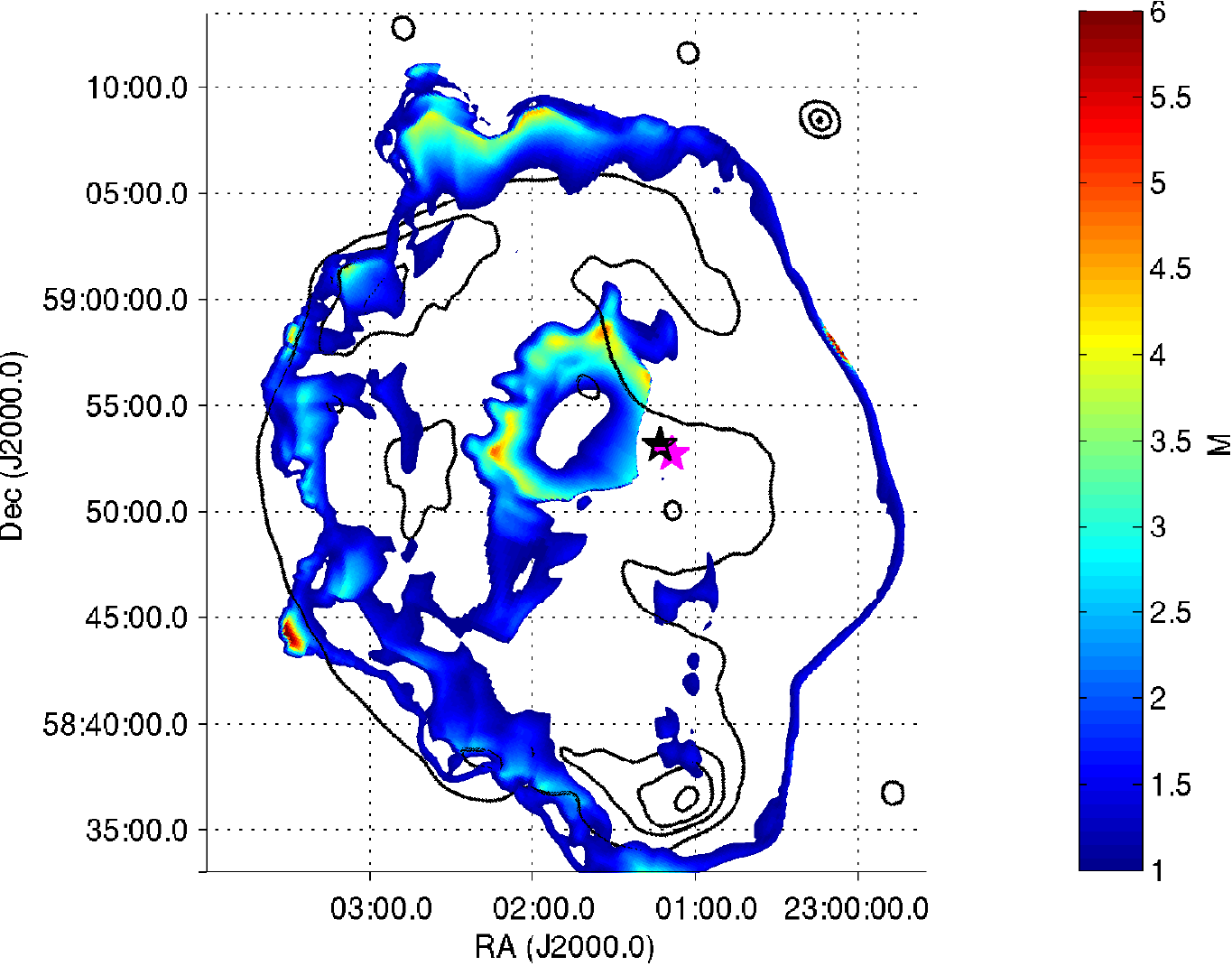}\hspace*{2mm}
  \includegraphics[width=\columnwidth]{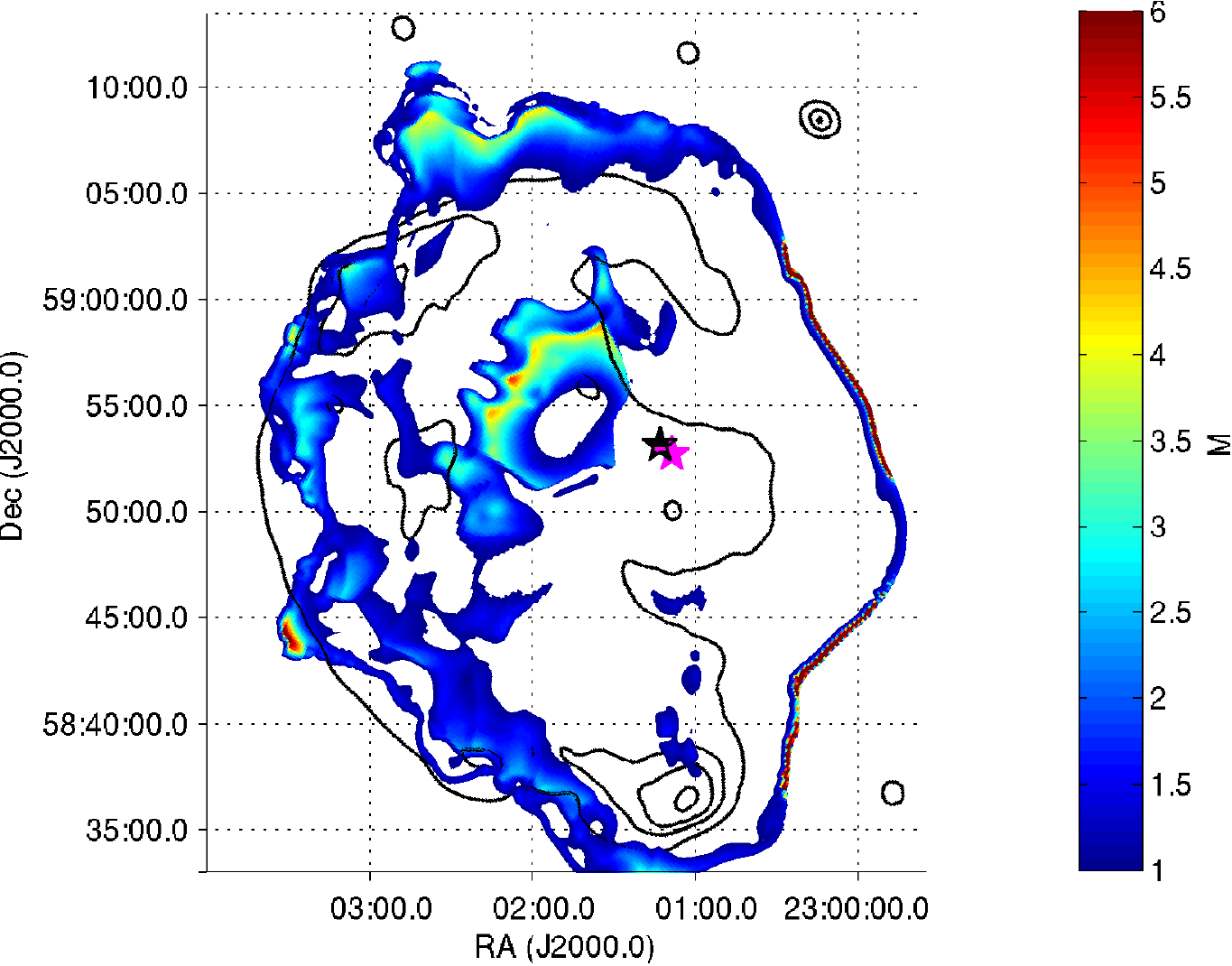}
\else
  \includegraphics[width=0.49\columnwidth]{radialvelocity.pdf}
  \includegraphics[width=0.49\columnwidth]{radialvelocity_chandra.pdf}\\
  \includegraphics[width=0.49\columnwidth]{machnumber.pdf}
  \includegraphics[width=0.49\columnwidth]{machnumber_chandra.pdf}
\fi
\makeatother
\caption{\emph{Left:}
Radial velocity from the explosion centre after 8,000~yr of our fiducial model based on the \emph{XMM-Newton} data (upper panel).  
Supersonic Mach numbers after 8,000~yr for our fiducial model based on the \emph{XMM-Newton} data (lower panel).  
\emph{Right:}
Radial velocity from the explosion centre after 11,000~yr of our fiducial model based the \emph{Chandra} data (upper panel). 
Supersonic Mach numbers after 11,000~yr for our fiducial model based on the \emph{Chandra} data (lower panel).
Regions with Mach numbers less than 1 are not displayed, hence they appear white. The position of the supernova explosion is marked with
a black star and the position of the AXP with a magenta star.}
\label{fig:velocity}
\end{figure*} 

In Fig.~\ref{fig:velocity} the computed radial 
velocity from the explosion centre and the supersonic Mach numbers of our 
fiducial models are displayed.  Regions with Mach numbers less than 1 are not displayed, hence they appear white.
The images in the left column show the results for which
we assumed the initial conditions derived from the \emph{XMM-Newton} data,
and the right column shows the results for the \emph{Chandra} case.

The average of the simulated shock velocities for the \emph{XMM-Newton}
case (Fig.~\ref{fig:velocity} (upper left)
is consistent with
the estimated mean shock velocity of $720\pm 60\,\text{km}\,\text{s}^{-1}$ by 
\citet{2004ApJ...617..322S}. In the region of the shock
the Mach number is $\sim 3$ 
(Fig.~\ref{fig:velocity} (lower left), 
which is at the limit of a strong shock. In the region of the shocked additional
cloud the Mach number exceeds the value of 5.

In particular, two regions with high Mach numbers can be identified. Firstly,
there is a ``blob'' in the eastern part where the shell breaks up and hot gas ($\sim 10^6~\text{K}$) escapes with
high velocity from the interior of the SNR. Secondly, the Mach number is higher in a shell-like structure where the dense Lobe-forming
cloud slowly dissolves.

The computed radial velocity from the explosion centre and the supersonic Mach numbers of our fiducial model
for the \emph{Chandra} data are displayed in Fig.~\ref{fig:velocity} (right). 
The mean value of the simulated shock velocities is again in agreement with the estimated mean shock velocity
of $460\pm 30\,\text{km}\,\text{s}^{-1}$ by \citet{2013A&A...552A..45S}. The value of the Mach number  
is again $\sim 3$ in the region of the shock and also exceeds the value of 5 where the additional
cloud was shocked. The same high Mach number structures as in the case of the \emph{XMM-Newton} data can also be seen here.

\section{Discussion}
Observations of the SNR CTB~109, e.g. by \citet{2004ApJ...617..322S,2006ApJ...642L.149S,2013A&A...552A..45S}, give
evidence that the bright diffuse X-ray emission in the remnant is the result of a shocked dense cloud. Our
hydrodynamic simulations support this hypothesis.

For our best fit model of the initial cloud properties we obtained a density of $n_\text{cloud}=20~\text{cm}^{-3}$
(\emph{XMM-Newton} data) and  $n_\text{cloud}=25~\text{cm}^{-3}$ (\emph{Chandra} data), respectively,
and a maximum size $d=4.54~\text{pc}$ for the shocked cloud. If we take into account that the ambient
ISM density derived from the \emph{XMM-Newton} and the \emph{Chandra} spectra have uncertainties between
10 to 60\% because of uncertainties  in the spectral fits and differences in the spectra for different
extraction regions, the cloud densities calculated for the \emph{XMM-Newton} and the \emph{Chandra} case are
consistent. The obtained values are much higher than the findings of the simple
analysis presented in \citet{2004ApJ...617..322S}, namely $n_\text{cloud}\lesssim5~\text{cm}^{-3}$ and $d \approx 1~\text{pc}$,
which was based on the analytic estimates given by \citet{1975ApJ...197..621S}. However, the density values are lower
than the value of \citet{2012ApJ...756...88C} from \emph{Fermi} observations, which is $n_\text{cloud}\approx 120~\text{cm}^{-3}$.

Our simulation reproduces for the preshock density $n_0 = 0.155~\text{cm}^{-3}$ (mixture of ISM and ejecta) from \emph{XMM-Newton} observations, an
initial energy of $E_0= 10^{51}~\text{erg}$, and the SN position $(\ell , b)=(109.0962^\circ , -0.9942^\circ)$
after 8,000~yr of evolution the observed morphology and size of CTB~109 (cf. Fig.~\ref{fig:density}). Similarly,  
the morphology of the remnant is matched very well after 11,000~yr for the higher value $n_0 = 0.3~\text{cm}^{-3}$ (ISM only) obtained
from the \emph{Chandra} data. 
These positions are consistent with the position of the SN that can be derived from the proper motions measurements for the
AXP by \citet{2013ApJ...772...31T}. The SN position, which can be calculated from the proper motion measurement by
\citet{2009AJ....137..354K} is rather inconsistent with our calculations.

The obtained age is comparable with the results from \citet{2004ApJ...617..322S} if the lower preshock density
value based on the \emph{XMM-Newton} data is taken. For $d_3=3.2/3.0$ they estimated an initial energy of
$E_0=(8.7\pm3.4)\cdot 10^{50}~\text{erg}$ and a remnant age of $t=(9.39\pm0.96)\cdot 10^{3}~\text{yr}$. An
analysis of the \emph{Chandra} observations gives a higher value of $t=(1.4\pm0.2)\cdot 10^4~\text{yr}$
\citep{2013A&A...552A..45S}, which was recently verified by \emph{Suzaku} observations \citep{2015PASJ...67....9N}.
This is comparable with the 11,000~yr, which we obtained from our simulations
with the \emph{Chandra} data, and also with the age of the remnant, which \citet{1992ApJ...388..127W} derived.
They have modelled CTB~109 numerically with the thin shell approximation and assumed
an initial energy of $E_0=3.6\cdot10^{50}~\text{erg}$, an ambient ISM density of $n_0=0.13~\text{cm}^{-3}$,
and a cloud density of $n_\text{cloud}=36~\text{cm}^{-3}$. After $t= 1.3 \cdot 10^4~\text{yr}$ their model
reproduced a semicircular shell of the observed size. In contrast to their thin shell approximation, our model
does not need a special geometry or symmetry assumptions, but solves the hydrodynamic equations for a completely inhomogeneous,
complex 3D realistic ISM with high resolution. Consequently, our model gives improved estimates on the
initial energy and the remnant age and can also explain its observed morphology.

In a similar manner we see that the shock velocities from the simulations and observations also agree
very well for the different preshock densities. For the simulation with the \emph{XMM-Newton} data the mean
shock velocity fits with $v_\text{s}=720\pm 60\,\text{km}\,\text{s}^{-1}$ \citep{2004ApJ...617..322S}
and the simulations with the \emph{Chandra} data agree very well with $v_\text{s}=460\pm 30\,\text{km}\,\text{s}^{-1}$
\citep{2013A&A...552A..45S}. The latter value of shock velocity was also obtained from the recent
\emph{Suzaku} observations \citep{2015PASJ...67....9N}.

A high synchrotron emission, which is visible in the radio continuum, correlates with more efficiently accelerated
electrons. Hence, higher shock velocities are expected in this region. The distribution of the shock velocities in
Fig.~\ref{fig:velocity} show high velocities in a shell-like
structure in the north-east and more concentrated in the south. These structures match quite well with the 1420-MHz
radio continuum of CTB~109, displayed in Fig.~\ref{fig:xmm} (left). This indicates that the simulated hydrodynamic structure
is a plausible representation of the remnant's structure.

At present, all parameter studies that we have performed give very good agreement with both \emph{XMM-Newton}
and \emph{Chandra} data sets. Therefore, we cannot favour one model over the other.
However, since the Chandra data
have been fitted with a more realistic two-component NEI model instead of a simplistic one-component NEI model,
we believe that the density model obtained from the \emph{Chandra} data is a better description of the reality.
Consequently, the obtained cloud properties, remnant age, and shock velocities from this model should be more reliable.

\section{Concluding remarks}
The presented numerical simulations of a supernova explosion in a realistic inhomogeneous medium
make it very plausible that the Lobe region is the result of a shocked dense elliptical cloud.
In the best fit initial density model the cloud has a density of $n_\text{cloud}=25~\text{cm}^{-3}$, which
lies between the values from the simple analysis by \citet{2004ApJ...617..322S} and the estimate by
\citet{2012ApJ...756...88C}. Not only the observed morphology of the remnant and of the Lobe, but also the
distribution of the observed X-ray intensity in the Lobe region could be reproduced very well by this model.
For an initial explosion energy of $E_0= 10^{51}~\text{erg}$ the remnant age can be estimated to be
11,000~yr. The result agrees with the remnant age of $\sim14,000~\text{yr}$ derived by \citet{2013A&A...552A..45S}.
This also allows us to derive the most probable site of the SN explosion.

Our simulations have shown that present observations in $^{12}$CO, \textsc{Hi}, and X-ray, constrain the
initial conditions well enough to derive important physical parameters, e.g. explosion energy and also
evolution time of an SNR. The simulations of the SNR CTB~109 show also that the preshock density distribution 
must have been more complex than simply a medium with a homogeneous density gradient.

Thus, we are confident that our method of using the present ambient density distribution of the SNR and modifying
it gradually to reproduce the morphology and X-ray emissivity of young SNRs by an ab initio 3D hydrodynamic simulation
is reasonable and will help to understand the realistic
conditions. 
This method will be explored in detail in forthcoming studies.

\begin{acknowledgements}
The research presented in this paper has used data from the Canadian 
Galactic Plane Survey, a Canadian project with international partners, 
supported by the Natural Sciences and Engineering Research Council. 

J.B. gratefully acknowledges support from the DFG via the Cluster of Excellence ``Origin and Structure of the Universe'' and J.B.
and D.B. acknowledge support from the priority program SPP 1573 ``Physics of the Interstellar Medium'' through the grant BR1101/7-1.

M.S. acknowledges support by the Deutsche Forschungsgemeinschaft through 
the Emmy Noether Research Grant SA2131/1-1.
\end{acknowledgements}
\bibliographystyle{aa} 
\bibliography{ctb109.bib}
\end{document}